\documentclass[final,3p,12pt]{elsarticle}
\usepackage{graphicx,amsbsy,amsmath,epsfig,natbib,float}
\usepackage{amssymb,latexsym,wasysym,pifont,mathrsfs}
\usepackage{comment,verbatim,multirow,array,sidecap,color}
\usepackage{lineno,setspace,geometry,soul,cancel}
\newfloat{widefig}{thp}{lop}
\journal{Physica A}

\begin{document}

\begin{frontmatter}

\title{Stochastic sandpile model on small-world networks: scaling and
  crossover}

\author{Himangsu Bhaumik}
\author{S. B. Santra}
\ead{santra@iitg.ac.in}
\address{Department of Physics, Indian Institute of Technology
  Guwahati, Guwahati-781039, Assam, India.}

\date{\today}
 
\begin{abstract}  
  A dissipative stochastic sandpile model is constructed on one and
  two dimensional small-world networks with different shortcut
  densities $\phi$ where $\phi=0$ and $1$ represent a regular lattice
  and a random network respectively. In the small-world regime
  ($2^{-12} \le \phi \le 0.1$), the critical behaviour of the model is
  explored studying different geometrical properties of the avalanches
  as a function of avalanche size $s$. For both the dimensions, three
  regions of $s$, separated by two crossover sizes $s_1$ and $s_2$
  ($s_1<s_2$), are identified analyzing the scaling behaviour of
  average height and area of the toppling surface associated with an
  avalanche. It is found that avalanches of size $s<s_1$ are compact
  and follow Manna scaling on the regular lattice whereas the
  avalanches with size $s>s_1$ are sparse as they are on network and
  follow mean-field scaling. Coexistence of different scaling forms in
  the small-world regime leads to violation of usual finite-size
  scaling, in contrary to the fact that the model follows the same on
  the regular lattice as well as on the random network
  independently. Simultaneous appearance of multiple scaling forms are
  characterized by developing a coexistence scaling theory. As SWN
  evolves from regular lattice to random network, a crossover from
  diffusive to super-diffusive nature of sand transport is observed
  and scaling forms of such crossover is developed and verified.
 
  \end{abstract}


\end{frontmatter}

\section{Introduction} 
The spontaneous emergence of spatio-temporal correlation and hence the
criticality in many natural phenomena are found to be outcome of self-organized criticality (SOC) \cite{bak,jensen,pruessner}. The
appearance of SOC on real-world complex networks is ubiquitous in
nature. Examples are spreading from Biology to Astrophysics to
technology such as, avalanche mode of activity in the neural network
of brain \cite{arcangelisPRL06,hesseFSN14}, earthquake dynamics on the
network of faults in the crust of the earth \cite{lisePRL02}, rapid
rearrangement of the magnetic field network in the corona
\cite{hughesPRL03}, propagation of information through a network with
a malfunctioning router causing the breakdown of the Internet network
\cite{motterPRE02}, breakdown of the electric power grid
\cite{carrerasIEEECS04} and many others.

On the other hand, small-world network (SWN) \cite{wattsNAT98} has a
very interesting property. It not only interpolates between the
regular lattice (RL) and the random network (RN) but also it preserves
the properties of both RL and RN, namely high
``clustering-coefficient''(concept of neighborhood) and
``small-world-effect'' (small average shortest distance between any
two nodes) respectively. It is always intriguing to study the models
of SOC on networks such as SWN. Sandpile, a prototypical model to
study SOC on RL introduced by Bak, Tang, and Wiesenfeld (BTW)
\cite{btwPRL87,btwPRA88} gives rise to anomalous (multi) scaling
\cite{deMenechPRE98,tebaldiPRL99,lubeckPRE00a}. The scaling behaviour
of such a deterministic model when studied on RN exhibited very
different behaviour than its behaviour on RL \cite{bonabeau95}. A Mean-Field
(MF) \cite{christensenPRE93,bonabeau95,gohPRL03,bhaumikPRE13} scaling
behaviour found to describe the avalanche properties rather than an
anomalous scaling of BTW model on RL. Stochastic sandpile models (SSM)
which incorporates random distribution of sand grains during
avalanche, exhibit a scaling behaviour with definite critical
exponents that follows finite-size scaling (FSS) when studied on
RL. SSM defines a robust universality class called Manna class
\cite{mannaJPA91,dharPHYA99a}. Not only SSM shows robust scaling
behaviour than the deterministic BTW model, it is also able to explain
certain experimentally observed avalanche behaviour
\cite{fretteNAT96}. SSM found to be one of the most studied models in
various dimensions in SOC literature
\cite{huynhPRE10,huynhJSM11,huynhPRE12}. However, there are not many
studies that report the scaling behaviour of SSM on SWN. It is then
important to study a stochastic sandpile model like SSM on SWN and
verify whether all such scaling forms would be preserved or not.

In this paper, a dissipative stochastic sandpile model (DSSM) is
developed and studied on SWN as a function of shortcut density $\phi$
in one and two dimensions. Dissipation of the sand occurs in the bulk
of the system in an annealed manner rather than at the
boundary. Various scaling forms of different avalanche properties are
identified and numerically verified.

\section{The model: DSSM on SWN} 
SWN is generated by adding shortcuts between two randomly chosen sites
both of a one dimensional ($1$d) lattice and those of a two
dimensional ($2$d) square lattice. The shortcut density, number of
added shortcuts per existing bond, is defined as $\phi=N_\phi/(dL^d)$
where $N_\phi$ is the number of shortcuts added and $dL^d$ is the
number of bonds (without shortcuts) present in a $d$-dimensional
lattice of linear size $L$ with periodic boundary conditions (pbc) in
both the directions. Care has been taken to avoid self-edges of any
node and multi-edges between any two nodes. The system corresponds to
a RL as $\phi\rightarrow 0$, whereas in the limit $\phi\rightarrow 1$
it corresponds to a RN. The system behaves like SWN for the
intermediate values of $\phi$ such as $2^{-12}$ to $2^{-3}$
\cite{bhaumikPRE13,newmanbook}. To study the sandpile dynamics on an
SWN, a suitable value of $\phi$ is chosen and the SWN is driven by
adding sand grains, one at a time, to randomly chosen nodes. If the
height $h_i$ of the sand column at the $i$th node becomes greater than
or equal to the predefined threshold value $h_c$, which is equal to
$2$ here, the $i$th node will topple and the height of the sand column
of the $i$th node will be reduced by $h_c$. The toppled two sand
grains are then distributed among two of its randomly selected
adjacent nodes which are connected to the toppled node either by
shortcuts or by nearest neighbour bonds. The toppling rule of the
$i$th critical node of DSSM can be written as
\begin{equation}
\label{3trule2}
\begin{array}{ll}
& h_i \rightarrow h_i-h_c, \\ {\rm and} & h_j=
  \left\{\begin{array}{ll} h_j  & \mbox{if} \hspace{0.2cm} r \le
  \epsilon_\phi, \\ h_j + 1 & \mbox{otherwise} \end{array}\right.
\end{array}
\end{equation}
where $j$ is two randomly and independently selected adjacent nodes,
and $r$ is a random number uniformly distributed between $0$ and $1$.
During transport of sand grains from one node to another, every sand
grain is attempted for dissipation from the system with a probability
$\epsilon_\phi$ to avoid overloading of the system.  If the toppling
of a node causes some of the adjacent nodes unstable, subsequent
toppling follow on these unstable nodes in a cascading manner which
leads to an avalanche. An avalanche stops when there is no unstable
node present in the system. During an avalanche no sand grain is added
to the system to ensure that the system in under slow driving force
and at the same time has a fast relaxation, which manifests the
separation of time scale, the key ingredient in a SOC system.

The probability of dissipation $\epsilon_\phi$ is taken as the inverse
of the average number of steps $\langle n_{\phi} \rangle$ required for
a random walker to reach the lattice boundary starting from any
lattice point \cite{malcaiPRE06,bhaumikPRE13}. The dissipation factor
$\epsilon_\phi=1/\langle n_{\phi} \rangle$ for a given $\phi$ is then
determined using the numerically estimated values of $\langle n_{\phi}
\rangle$.

\section{Numerical simulations and steady state}
\begin{figure}[t]
\centerline{\hfill
  \psfig{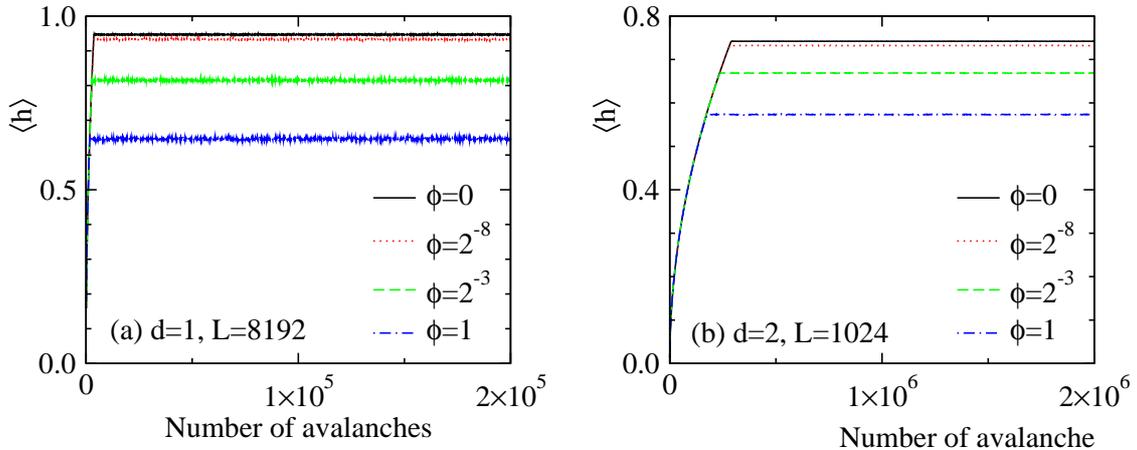}\hfill
}
\caption{ Plot of $\langle h \rangle$ against the number of avalanches
  for different values of $\phi: 0$ (solid black line) , $2^{-8}$
  (dotted red line), $2^{-3}$ (dashed green line), and $1$ (blue
  dotted dashed line) for $1$d with $L=8192$ in (a), and for $2$d with
  $L=1024$ in (b). }
\label{3avgh}
\end{figure}

Extensive computer simulations are performed to study the dynamics of
DSSM on SWN for various values of $\phi$ and system sizes $L$ both in
$1$d and $2$d. For $1$d, $L$ is varied from $1024$ to $8192$, whereas
for $2$d, $L$ is varied from $128$ to $1024$ in multiple of $2$. To
estimate $\epsilon_\phi$, for a given $L$ and $\phi$, $\langle n_\phi
\rangle$ is calculated by performing $2 \times 10^6 $ random walks in
$16$ different random configurations of SWN. Knowing $\epsilon_\phi$,
starting from the empty configuration of $h_i$ (i.e., $h_i=0 \ \forall
\ i$), sand grains are added at random positions. The average height
of the sand column $\langle h \rangle =\sum_0^{L^d}h_i/L^d$ for three
different values of $\phi$ are plotted against number of avalanches in
Figs. \ref{3avgh}(a) and \ref{3avgh}(b) for $1$d and $2$d
respectively. It can be observed that after a transient period the
system evolves to a steady state which corresponds to equality of
current of sand influx (due to adding sand) and current of sand
outflux (due to dissipation of sand). Such balance of sand influx and
outflux maintains the system in a critical steady state. Critical
properties of DSSM on SWN are characterized studying various
properties of avalanche like size $s$ (total number of toppling in an
avalanche), area $a$ (number of distinct nodes toppled in an
avalanche), etc., in the steady state at different values of shortcut
density $\phi$. For a given $L$ and $\phi$, data are averaged over
$32\times 10^6$ avalanches (collected in the steady state) on $32$
different SWN configurations. The information of an avalanche is kept
by storing the number of toppling of every node in an array
$S_\phi[i],i=1,\cdots,L^d$ which was set to zero initially. All
geometrical properties of an avalanche such as avalanche size $s$,
avalanche area $a$, etc., can be estimated in terms of $S_\phi[i]$ as
\begin{equation}
s=\sum_{i=1}^{L^d} S_\phi[i], \ \ \ a= \sum_{i=1}^{L^d}1
\end{equation}
for all $S_\phi[i]\ne 0$.

\section{Results and discussion}
The critical properties of stochastic sandpile model are already known
on RL ($\phi=0$) \cite{lubeckPRE00a,huynhJSM11,huynhPRE12} as well as
on RN ($\phi=1$) \cite{moosaviPRE14,bhaumikPRE16}. In the present
study, the critical behaviour of DSSM on small-world regime
($2^{-12}\le \phi \le 0.1$) will be addressed. Limiting behaviour of
such result would confirm the results corresponding to $\phi=0$ and
$\phi=1$.

\subsection{Toppling surface: fragmentation, compactness, and fluctuation}
\begin{figure}[t]
\centerline{\hfill
  \psfig{file=figure2abc.eps,width=0.9\textwidth}\hfill
}
\vspace{0.4cm}
\centerline{\hfill\hfill\hfill\hfill
\psfig{file=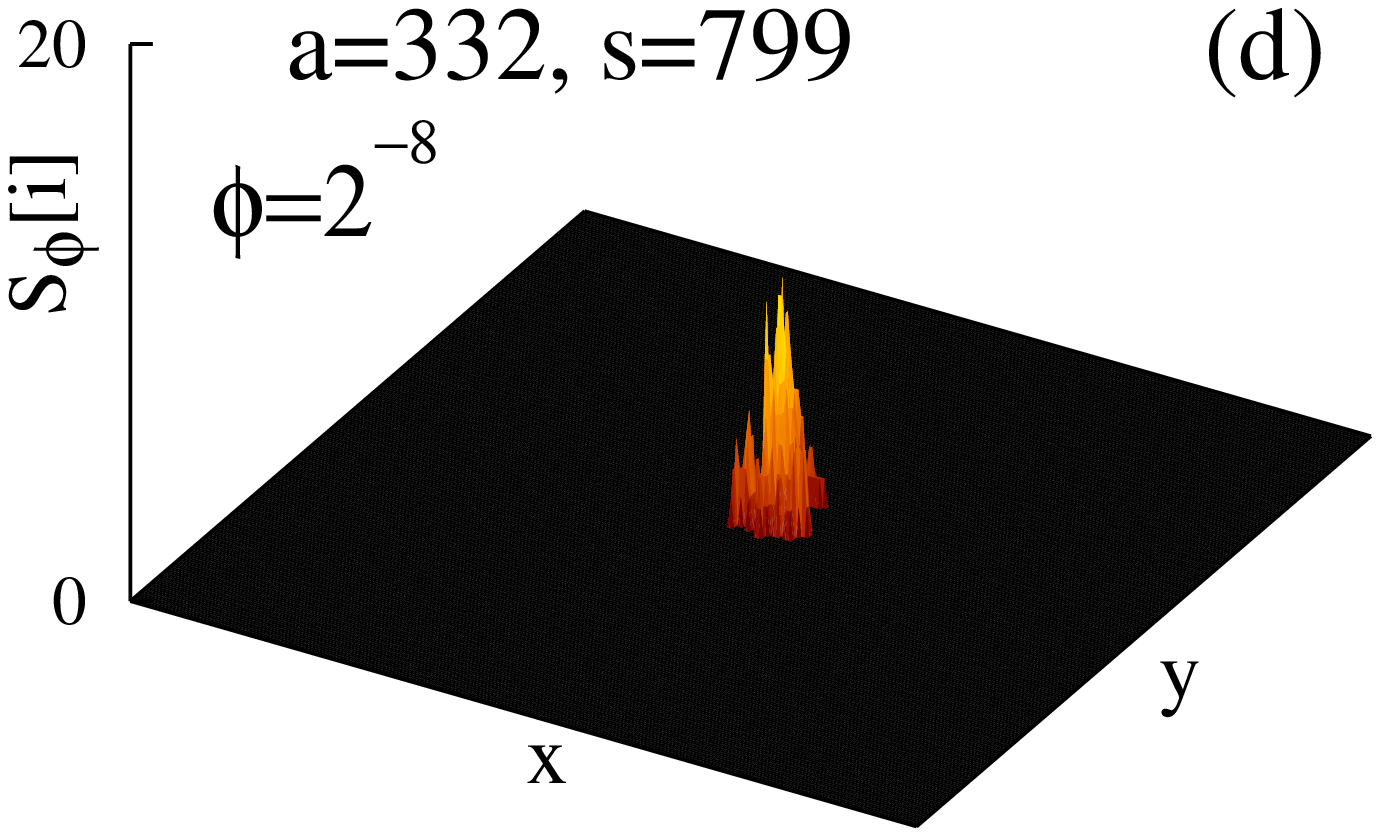,width=0.29\textwidth}\hfill
\psfig{file=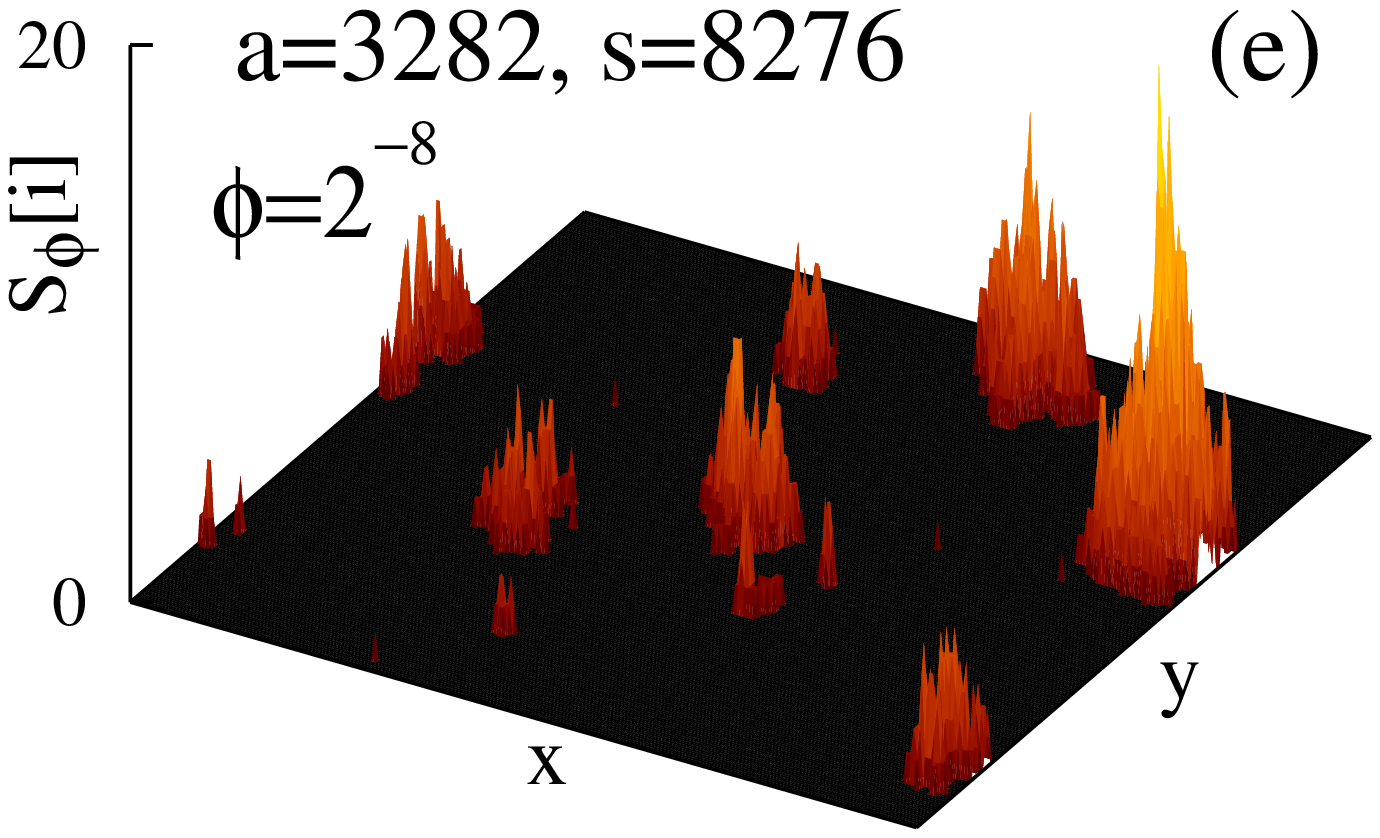,width=0.29\textwidth}\hfill
\psfig{file=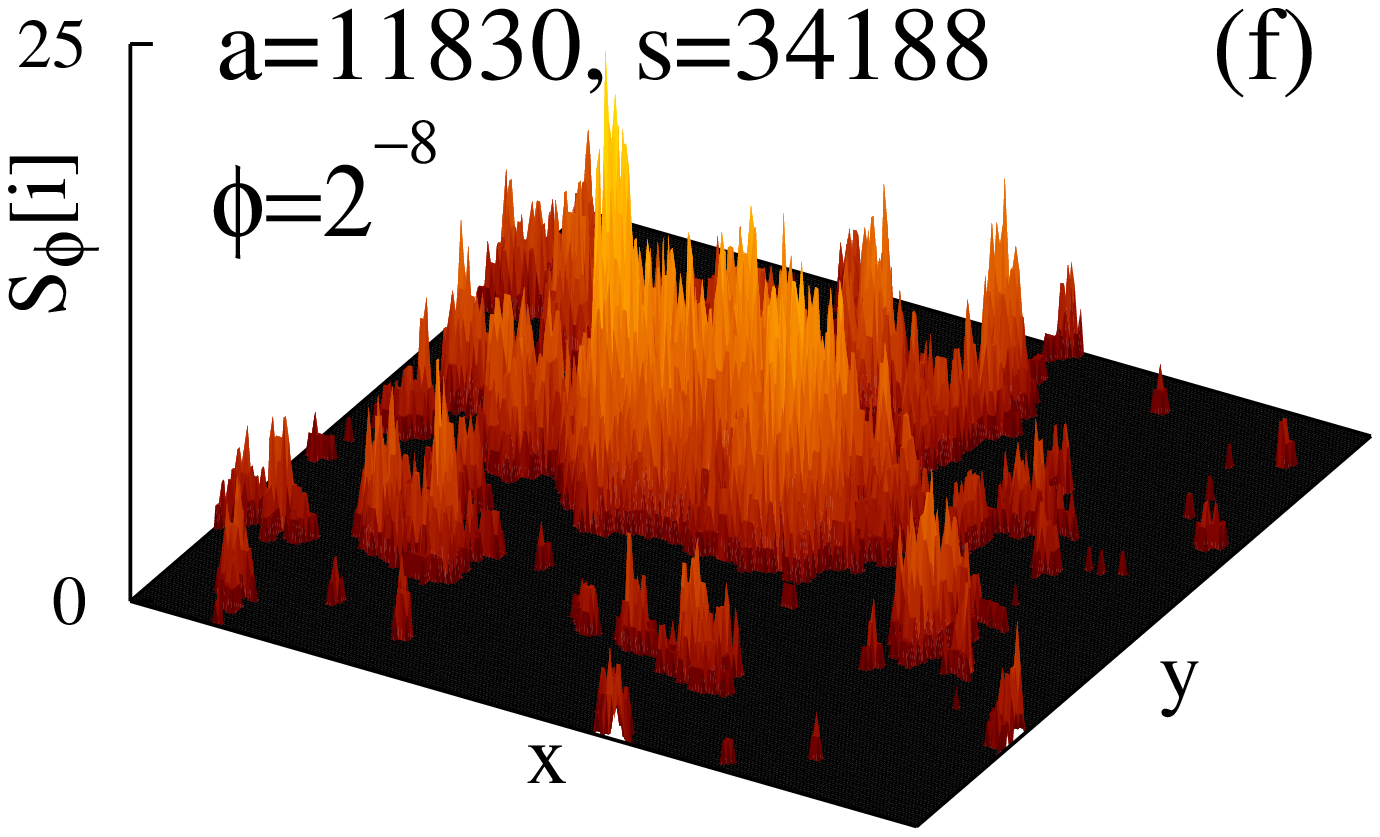,width=0.29\textwidth}\hfill\hfill\hfill\hfill
}
\caption{ Toppling surfaces of various avalanche clusters of different
  avalanche area ($a$) and sizes ($s$) in the SWN regime are
  shown. Toppling surfaces on $1$d lattice of $L=256$ and
  $\phi=2^{-6}$ are presented in the upper panel and those on $2$d
  square lattice of size $L=256$ and $\phi=2^{-8}$ are presented in
  the lower panel. The values of $s$ and $a$ are mentioned as legends
  in the respective plots. }
\label{3surf}
\end{figure}

In order to characterize various geometrical properties of avalanche
one needs to visualize the avalanche in a suitable parameter space.
The values of the toppling number $S_\phi[i]$ of an avalanche at
different nodes of SWN define a surface called toppling surface
\cite{ahmedEPL10} which not only serves as an important quantity to
visualize an avalanche but also presents important scaling behaviour
of several geometrical properties of the avalanche
\cite{ahmedPHYA12,bhaumikPRE14}. For an intermediate value of $\phi$
(SWN regime), the toppling surfaces of DSSM for both $1$d and $2$d are
presented in Fig. \ref{3surf} for various avalanches of different area
and sizes. The upper panel corresponds to the toppling surfaces on
$1$d for $L=256$ and $\phi=2^{-6}$, and the lower panel represents
those on a $2$d square lattice of size $L=256$ and
$\phi=2^{-8}$. Since the avalanche clusters are occurring on a
network, it might be sparse in the sense that different parts of the
same cluster are separated by an Euclidean distance greater than the
lattice spacing. A cluster is said to be compact if all the toppled
sites are only separated by nearest neighbour lattice
spacing. Otherwise the cluster will be called sparse or fragmented. It
can be seen that for both $1$d and $2$d, the avalanches of smaller
area are compact as those are confined in a length scale where SWN
behaves as a RL. The avalanches of intermediate area and sizes are
found to be sparse as those avalanches are exposed to the network. The
avalanches of area comparable to system size in $1$d are also found
compact (see Fig.\ref{3surf}(b)) whereas those in $2$d still remain
sparse (see Fig.\ref{3surf}(f)). Hence, not only the characteristic
behaviour of avalanches are different for different values of
avalanche size $s$, but also they differ considerably on different
dimensions.

\begin{figure}[t]
\centerline{\hfill 
  \psfig{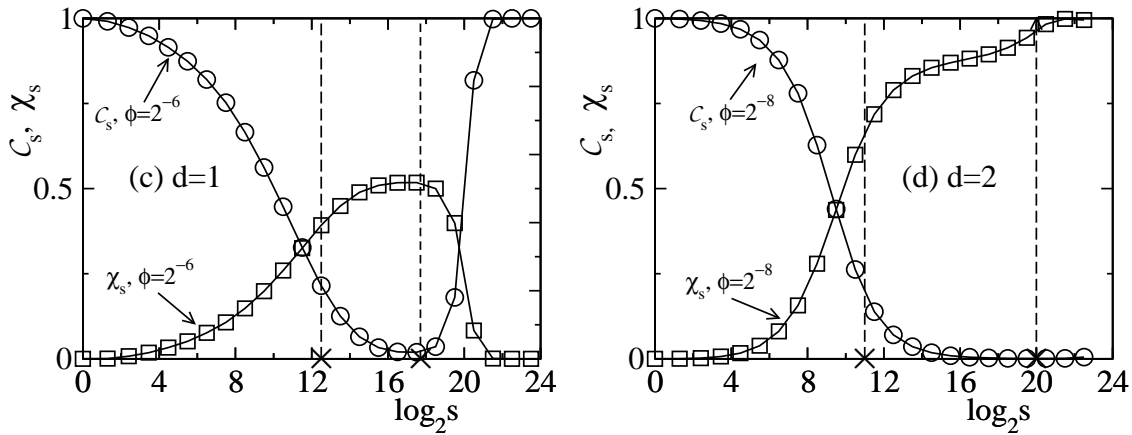}\hfill
}
\caption{ Plot of $\mathcal {C}_s$ ($\Circle$) and $\chi_s$ ($\Box$)
  against $s$ in semi-logarithmic scale for $1$d in (a) and for $2$d
  in (b). Data for $1$d are collected for $\phi=2^{-6}$ on the system
  size $L=8192$, whereas, that for $2$d are collected for
  $\phi=2^{-8}$ and $L=1024$. Vertical dashed lines in each figure
  represent tentatively crossover sizes .}
\label{3CsXs}
\end{figure}

As the avalanches are compact single cluster with no fragmentation on
RL and are fragmented in several sub-cluster on RN (see
Fig. \ref{3surf}), the compactness of an avalanche need to be
studied. If an avalanche of size $s$ is fragmented into $N_f$ number
of fragments with area $a_f$ of the $f$th fragment, then the
compactness $\mathcal{C}_s$ and the fluctuation in fragment area,
$\chi_s$, can be defined as
\begin{equation}
\mathcal{C}_s=\left \langle\frac{1}{N_f}\right\rangle_s,
\ \ \ \ \chi_s=\left \langle \left[ 1- \frac{\langle a_f
    \rangle^2}{\langle a_f^2 \rangle }\right] \right\rangle_s
\end{equation}
giving $\mathcal{C}_s=1$ and $\chi_s=0$ for $N_f=1$ (no fragmentation)
and $\mathcal{C}_s \rightarrow 0$ and $\chi_s\rightarrow 1$ for large
$N_f$. For SWN regime, $\mathcal {C}_s$ and $\chi_s$ are plotted in
Figs. \ref{3CsXs}(c) and \ref{3CsXs}(d) for $1$d and $2$d
respectively. It can be seen that there exists three different regimes
of avalanche size $s$ in $1$d and two such regimes in $2$d where both
$\mathcal {C}_s$ and $\chi_s$ display different characteristic
behaviour with $s$. Such regimes are identified by the dashed lines at
the tentative crossover sizes $s_1$ and $s_2$, marked by crosses in
Fig. \ref{3CsXs}, which will be estimated later. Both in $1$d and
$2$d, $\mathcal{C}_s$ in the region $s_1<s<s_2$, is quite low, less
than $0.20$. Hence an avalanche cluster on an average is fragmented at
least into five pieces. These clusters are called sparse. Similarly,
$\chi_s$ attains a high value and remains more or less constant till
$s\approx s_2$. For the avalanches of size $s<s_1$, $\mathcal{C}_s$ is
high between $1$ and $0.20$ indicating a compact
avalanche. Consequently, $\chi_s$ is small. Whereas for the avalanches
of size $s>s_2$, the behaviour of $\mathcal{C}_s$ and $\chi_s$ are
very different in $1$d and $2$d. For $1$d, $\mathcal{C}_s$ starts
increasing with $s$ and attains $1$ and at the same time $\chi_s$ goes
to zero. The avalanches then become a single compact avalanche again
(see, Fig.\ref{3surf}(c)) whereas, in $2$d, $\mathcal{C}_s$ is close
to zero and $\chi_s$ is almost one. It suggests that in $2$d
avalanches are not only fragmented but also the fragments have high
fluctuation in their masses. These are consistence with the toppling
surface presented in Fig. \ref{3surf}(f). Usually on RL ($\phi=0$),
the avalanches are trivially found to be compact but on the random
network ($\phi=1$), they are found to be fragmented.

\subsection{Conditional expectation and scaling}
In order to characterize the toppling surface quantitatively, average
height $S_s$ and area $\langle a_s\rangle$ of the toppling surfaces
are studied as a function of the avalanche size $s$. The average area
$\langle a_s\rangle$ of an avalanche of fixed size $s$ is defined as
\begin{equation}
\label{3as}
\langle a_s\rangle = \int aP(a|s)ds 
\end{equation}
where $P(a|s)$ is the conditional probability for an avalanche of
area $a$ and size $s$ to appear \cite{christensenPRE93}. The average
height $\langle S_s\rangle$ of a toppling surface is defined as,
\begin{equation}
S_s= \frac{1}{a}\sum_{i=1}^{a} S_\phi[i]  =
\frac{s}{a},
\label{3Sse}
\end{equation}
which is then averaged over different avalanches of given size $s$ for
a given $\phi$. The scaling of $\langle a_s\rangle$ and $\langle
S_s\rangle$ with $s$ is given by
\begin{equation}
\label{3ass}
\langle a_s\rangle \sim s^{\gamma_{as}} \ \ \ {\rm and} \ \ \ \langle
S_s\rangle  \sim s^{1-\gamma_{as}}
\end{equation}
where $\gamma_{as}$ is an exponent.

\begin{figure}[t]
  \centerline{\hfill 
 \psfig{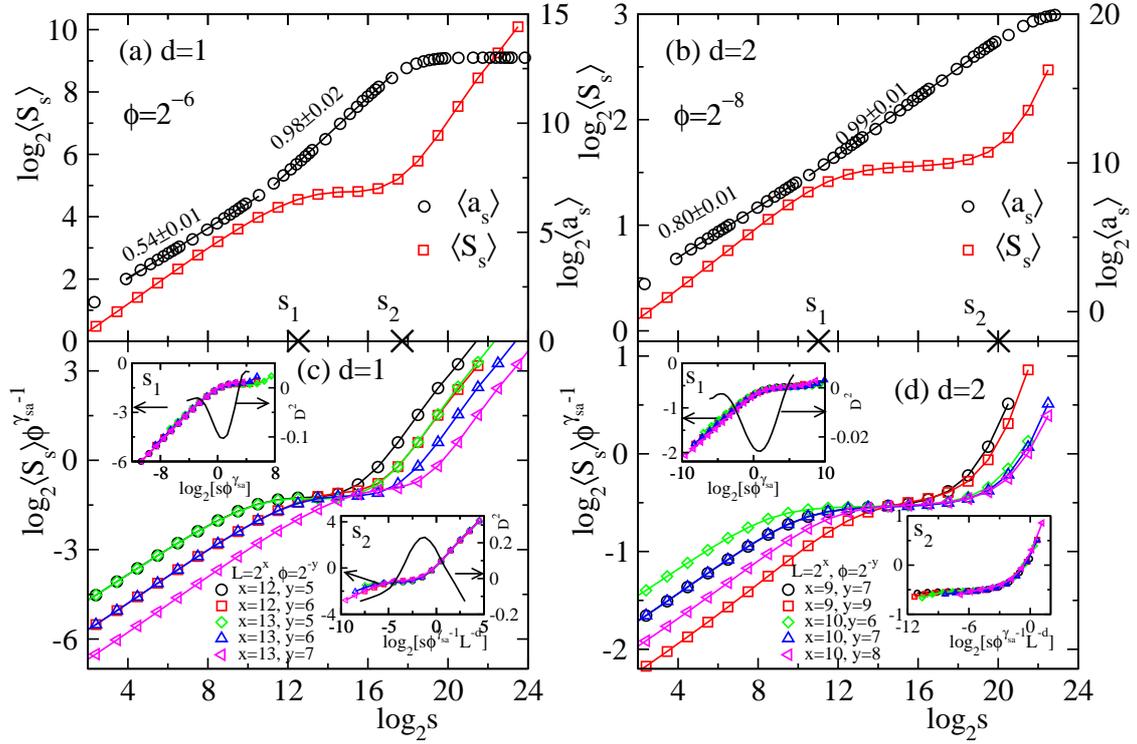}\hfill}
\caption{ Plot of $\langle a_s \rangle$ against $s$ (black circle) and
  $\langle S_s\rangle$ against $s$ (red square) for $1$d in (a) and
  for $2$d in (b). Data for $1$d are collected for $\phi=2^{-6}$ and
  $L=8192$, whereas, that for $2$d are collected for $\phi=2^{-8}$ and
  $L=1024$. The rescaled $\langle S_s\rangle$, $\langle S_s\rangle
  \phi^{\gamma_{sa}-1}$ is plotted against $s$ for various values of
  $\phi$ and $L$ (see legend) in (c) for $1$d and in (d) for
  $2$d. Insets: $\langle S_s \rangle\phi^{\gamma_{sa}-1}$ is plotted
  is plotted against scaled $s_1$ and $s_2$. Solid curve in each inset
  represents the double derivative of the scaled function against its
  argument.}
\label{3S}
\end{figure}

In the SWN regime, for a given system size $L$ and shortcut density
$\phi$, the estimated values of $S_s$ and $\langle a_s\rangle$ are
plotted against $s$ in Figs. \ref{3S}(a) and \ref{3S}(b) for $1$d and
$2$d respectively. The values of $\gamma_{as}$ are measured by linear
least square fit through the data points of $\langle a_s\rangle$ for
different region of $s$. For the avalanches of size $s<s_1$, it is
found that $\gamma_{as}=0.54\pm 0.01$ in $1$d and $0.80\pm 0.01$ in
$2$d in close agreement with the known values of $\gamma_{as}$ for SSM
on RL as $\gamma_{as}=1/2$ \cite{nakanishiPRE97} in $1$d and
$\gamma_{as}\approx 0.78$ \cite{benhurPRE96} in $2$d. The scaling
exponents of $\langle S_s\rangle$ are found as $\approx 0.5$ in $1$d
and $0.22$ in $2$d which is consistent with $1-\gamma_{as}$. The
scaling behaviour in this region is thus governed by the properties of
the avalanches on RL ($\phi=0$). For the avalanches of size $s>s_2$,
power law scaling of $\langle S_s\rangle$ and $\langle a_s\rangle$ are
observed only for $1$d and no such scaling behaviour are observed for
$2$d. In $1$d, $\langle a_s\rangle$ saturates in this region though
the avalanche size $s$ is increasing indicating $\gamma_{as}=0$ and
$1-\gamma_{as}=1$. For the avalanches of intermediate sizes, $s_1< s<
s_2$, $\gamma_{as}\approx 1$, ($0.98$) in $1$d and ($0.99$) in $2$d,
as shown in Figs. \ref{3S}(a) and \ref{3S}(b),
respectively. Accordingly, $\langle S_s\rangle$ remains constant
against $s$. Since $\gamma_{as} =1$ on RN in both $1$d and $2$d
\cite{christensenPRE93}, the scaling behaviour in this region is then
governed by the properties of the avalanches on random network. The
values of $s_1$ and $s_2$ can be obtained determining the points of
inflections in the plots of $\langle S_s\rangle$ against $s$. These
plots for different $\phi$ and $L$ are found to be shifted both
vertically and horizontally. A scaling form for $\langle S_s\rangle$
as well as for the crossover sizes $s_1$ and $s_2$ can be established
considering the fact that there exits a characteristic length $\xi\sim
\phi^{-1/d}$ where $d$ is the dimensionality of the lattice, below
which SWN belongs to ``large world'', the RL regime and beyond which
it behaves as ``small-world'', the RN regime
\cite{newmanPRE99,newmanPLA99,mendesEPL00}. At this length scale, the
avalanche area $a \approx \xi^d \sim 1/\phi$ and $\langle S_s\rangle$
can accordingly be obtained as
\begin{equation}
\label{3Ssphi}
\langle S_s\rangle =\frac{s}{a} \sim \phi^{1-\gamma_{sa}}
\end{equation}
where $\gamma_{sa}=1/\gamma_{as}$. In Figs. \ref{3S}(c) and
\ref{3S}(d), the scaled average height $\langle
S_s\rangle\phi^{\gamma_{sa}-1}$ are plotted against $s$ for $1$d and
$2$d respectively for different values of $L$ and $\phi$. It can be
seen that in the intermediate range of $s$, the plots in both the
figures have a common constant height independent of $\phi$ and
$L$. The avalanches confined within $\xi$ are expected to have $s<s_1$
whereas the avalanches extend beyond $\xi$ and comparable to $L$ are
expected to have $s>s_2$. Hence, two crossover sizes $s_1$ and $s_2$
correspond to two length scales, $\xi$ and $L$, present in this
system. The scaling form of $s_1$ and $s_2$, the crossover avalanche
sizes, are then given by
\begin{equation}
\label{3s1s2}
s_1= \bar{S}\xi^d \sim \phi^{-\gamma_{sa}} \ \ \ {\rm and}
\ \ \ s_2= \bar{S}L^d \sim \phi^{1-\gamma_{sa}}L^d
\end{equation}
where $\bar{S}$ is the value of $\langle S_s\rangle$ in the
intermediate region. Each data set of $\langle S_s\rangle$ is
subdivided into two sets breaking it at the middle of the intermediate
region. The scaling forms of $s_1$ and $s_2$ are verified
independently by data collapse for $\langle S_s\rangle$ in different
regions as shown in the insets of Figs. \ref{3S}(c) and
\ref{3S}(d). Identifying the point of inflection in the collapsed
plots (taking double derivative $D^2$ with respect to the argument),
the values of $s_1$ and $s_2$ can be estimated. The positions of the
dips and peaks of $D^2$ corresponding to the values of $s_1$ and $s_2$
are shown in the respective insets of Figs. \ref{3S}(c) and
\ref{3S}(d). However, the positions of the dips and peaks are found
slightly shifted from the origin in the logarithmic scale due to
associated metric factors in $s_1$ and $s_2$
(Eq. \ref{3s1s2}). Estimating the corresponding matric factors, the
values of $s_1$ and $s_2$ in $1$d for $L=2^{13}$ and $\phi=2^{-6}$ are
obtained as: $s_1\approx 2^{12.5}$ and $s_2\approx 2^{17.7}$. In $2$d,
however, no inflection point is found for $s_2$. Therefore, for
$L=2^{10}$ and $\phi=2^{-8}$, the value of $s_1$ is obtained
incorporating the metric factor as $s_1=2^{10.97}$ and that of $s_2$
by direct estimation from Eq. (\ref{3s1s2}) as $s_2=2^{20}$. The
values of $s_1$ and $s_2$ are marked by crosses in Figs. \ref{3S}(a)
and \ref{3S}(b). It has been verified that $\langle S_s\rangle$ of
deterministic BTW-type sandpile also exhibits similar scaling
behaviour at different regions of $s$ in the SWN regime.

The ensemble of avalanches that appear in the steady states of these
systems then can be classified into three different categories. The
avalanches with size $s<s_1$ are mostly fragment less and confined on
the large world or RL. The growth of their average height follow
different power law scaling with the avalanche size $s$ in $1$d and
$2$d. In the intermediate region $s_1<s<s_2$, the avalanche clusters
are sparse, different parts of the same cluster are connected by the
links of the network. These clusters are then appearing on the
small-world or on the network. Their average height do not grow with
their avalanche size as the avalanche cluster become sparse both in
$1$d and $2$d. As the size of the avalanches exceed $s_2$, the
avalanche cluster properties are very different in one and two
dimensions. In this region, the avalanche clusters become a compact
over grown single cluster in $1$d whereas in $2$d they grow in size
but remain sparse.

\subsection{Avalanche size and area distributions}

\begin{figure}[t]
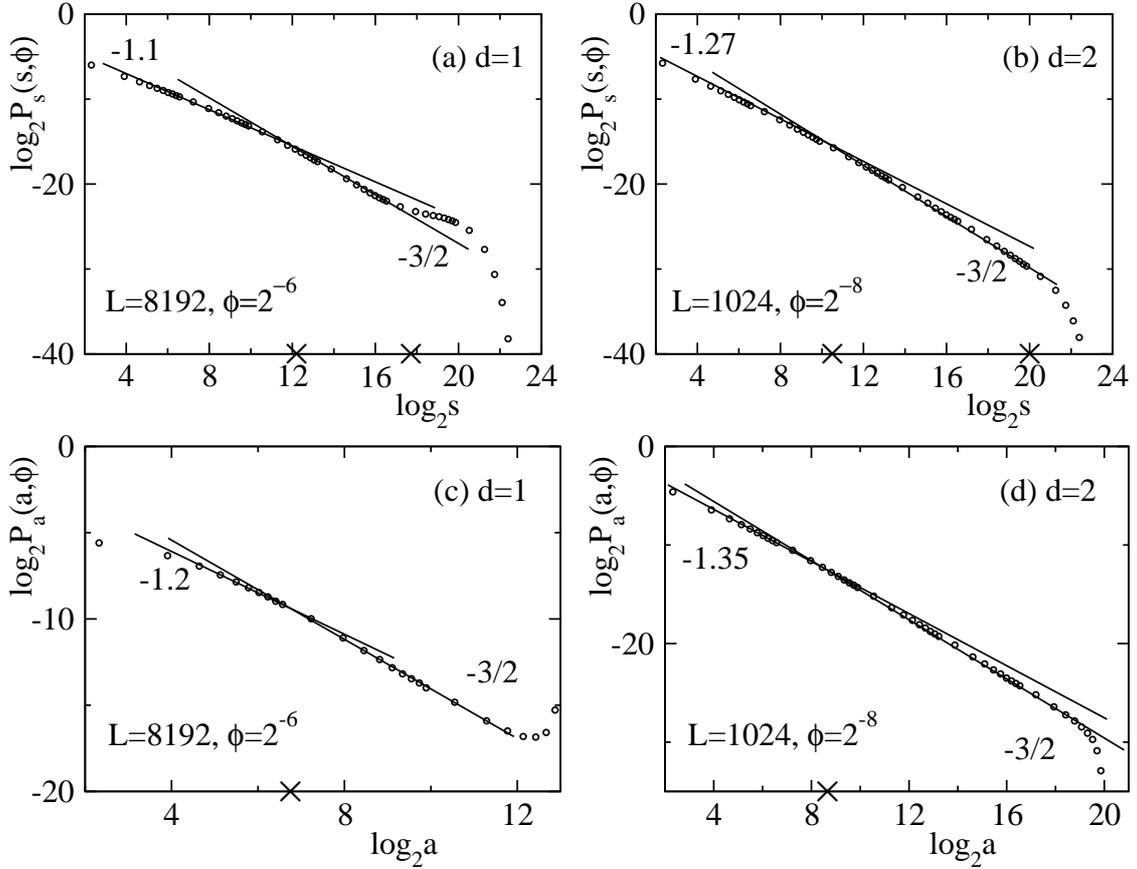

\centerline{\hfill
  \psfig{file=figure5ab.eps,width=0.9\textwidth}\hfill
} \centerline{\hfill
  \psfig{file=figure5cd.eps,width=0.9\textwidth}\hfill
}
\caption{ Plot of $P_s(s,\phi)$ against $s$ in (a) for $1$d with
  $L=8192$ for $\phi=2^{-6}$ and in (b) for $2$d, $L=1024$ for
  $\phi=2^{-8}$. Plot of $P_a(a,\phi)$ against $a$ in (c) for $1$d and
  (d) for $2$d with the same $L$ and $\phi$ as in (a) and (b)
  respectively. The solid lines with required slope (as indicated
  beside) represent two distinct scaling behaviour in two regimes.}
\label{3Ps1}
\end{figure}
The exponents associated with the power-law scaling of the probability
distributions of avalanche property $x\in \{s,a\}$ are generally use
to characterize the critical steady state of sandpile model. For a
given $\phi$, the probability to have an avalanche of property $x$ is
given by $P_x(x,\phi) = N_{x,\phi}/N_{\rm tot}$ where $N_{x,\phi}$ is
the number of avalanches having property $x$ out of total number of
avalanches $N_{\rm tot}$ generated at the steady state. Since the
avalanches occurring on a given SWN (of particular $\phi$) exhibits
different geometrical properties according to their size $s$, the
avalanche size distribution $P_s(s,\phi)$ should exhibit power-law
scaling with different exponents in different regimes in a single
distribution. A similar scaling behaviour of avalanche area
distribution $P_a(s,\phi)$ is also expected, because for each
crossover size there should a associated crossover area. A generalized
scaling form of $P_x(x,\phi)$ among any two regimes is proposed as
\begin{equation}
\label{3scaling}
P_x(x,\phi) = \left \{ \begin{array}{ll} x^{-\tau_{x1}} {\sf
    f_x}\left(\frac{x}{x_c(\phi)}\right) & \mbox{for $x \leqslant x_c$}
  \\ x^{-\tau_{x2}} {\sf
    g_x}\left(\frac{x}{x_c(\phi)}\right) & \mbox{for
    $x \geqslant x_c$ }
       \end{array} \right. 
\end{equation}
where $x_c$ is the crossover value of the property $x$, ${\sf f_x}$
and ${\sf g_x}$ are the respective scaling functions and $\tau_{x1}$,
$\tau_{x2}$ are the corresponding critical exponents in the respective
region. $P_x(x,\phi)$ are estimated on a $1$d lattice of size $L=8192$
for $\phi=2^{-6}$, and on a $2$d square lattice of size $L=1024$ for
$\phi=2^{-8}$. The distributions of avalanche size $P_s(s,\phi)$ are
presented in Figs. \ref{3Ps1}(a) and \ref{3Ps1}(b) respectively for
$1$d and $2$d lattices.  It can be seen that the distributions
$P_s(s,\phi)$ in Figs. \ref{3Ps1}(a) and \ref{3Ps1}(b) do not follow a
single power law scaling over the whole range of avalanche
sizes. There are three distinct regions of avalanche size $s$,
separated by two crossover sizes $s_1$ and $s_2$, which were observed
in case of geometrical aspect of the avalanches, indicated by
crosses. $P_s(s,\phi)$ seems to have different scaling exponents in
different regions of $s$. It could be observed that the avalanches of
size $s>s_2$ are large avalanches and do not contribute to the
power-law scaling. Hence, the crossover at $s_1$ and the associated
scaling properties will be analyzed in the following. Since the
avalanches in the region $s<s_1$ correspond to the avalanches on RL,
the scaling behaviour of $P_s(s,\phi)$ should either be that of Manna
scaling. Whereas, for the region $s>s_1$ the avalanches correspond to
those on the network, and hence the scaling behaviour should follow MF
scaling on network. The avalanche size distribution exponent $\tau_s$
of SSM on a regular $1$d lattice is known to be $1.1$
\cite{dickmanPRE03,nakanishiPRE97,dickmanBJP00,bonachelaPRE08,
  bhaumikPRE16} and that on a regular $2$d lattice is $\approx 1.28$
\cite{lubeckPRE00a,bhaumikPRE16}. Whereas the MF value of $\tau_s$ is
$3/2$ on both $1$d and $2$d. Two straight lines of respective slopes
are plotted in Figs. \ref{3Ps1}(a) and \ref{3Ps1}(b) as guide to the
eye. It can be noticed that a reasonable portion of data points of
$P_s(s,\phi)$ obtained for SWN do follow the respective scaling forms
in both the dimensions. The avalanche area distributions $P_a(a,\phi)$
are presented in Figs. \ref{3Ps1}(c) and \ref{3Ps1}(d) respectively
for $1$d and $2$d lattices. A similar behaviour as that of
$P_s(s,\phi)$ is observed for $P_a(a,\phi)$ too. The value of
crossover area $a_1$ associated with crossover size $s_1$ can be
estimated from the conditional expectation relation $a_1\sim
s_1^{\gamma_{as}}$. Taking the respective values of $s_1$ and
$\gamma_{as}$, the values of $a_1$ are estimated as $2^{6.75}$ for
$1$d and $2^{8.77}$ for $2$d and are indicated by crosses in
Figs. \ref{3Ps1}(c) and \ref{3Ps1}(d) respectively. The avalanche area
distribution exponent $\tau_a$ of SSM on a regular $1$d lattice is
known to be $1.2$ and that on a regular $2$d lattice is $\approx 1.35$
\cite{huynhJSM11}, whereas the MF value is $\tau_a=3/2$. It can be
seen that data points of $P_a(a,\phi)$ follow SSM scaling in $a<a_1$
regime and MF scaling in $a>a_1$ regime in both the dimensions.

\begin{figure}[t]
\centerline{\hfill 
  \psfig{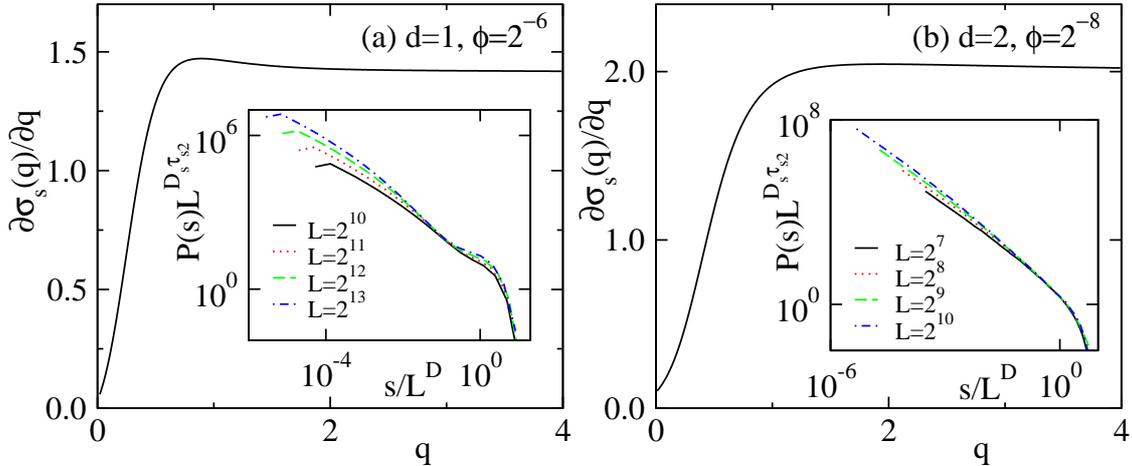}\hfill
 }
\caption{ Plot of $\partial \sigma_s(q,\phi)/\partial q$ against $q$
  in (a) for $1$d with $\phi=2^{-6}$ and in (b) for $2$d with
  $\phi=2^{-8}$. Inset: Attempt to collapse the data of avalanche size
  distribution for different system sizes taking $\tau_s=3/2$, and
  respective values of $D_s$. }
\label{3fss}
\end{figure}

It has already been known that the distribution functions of
stochastic sandpile model follow FSS both on RL ($\phi=0$) as well as
on RN ($\phi=1$) \cite{tebaldiPRL99,moosaviPRE14}. It is
then important to verify whether the probability distributions of DSSM
follow FSS or not in the SWN regime. The FSS scaling form of size
distribution function for a given fixed $\phi$ is assumed as
\begin{equation}
\label{3PsL}
P(s,L)\approx s^{-\tau}f(s/L^{D_s})
\end{equation}
 where $D_s$ is the capacity dimension. Moment
analysis of the avalanche size has been performed for both the
dimensions. The average $q$th moment of avalanche size $s$ for a given
$\phi$ is defined as
\begin{equation}
\label{3qmnts}
\langle s^q \rangle=\int_0^{s_{max}}s^qP(s,L)ds \sim
L^{\sigma_s(q)}
\end{equation}
where $\sigma_s(q)=[q+1-\tau_s]D_s$ is the moment scaling
function. The values of $\sigma_s(q)$ for different values of $\phi$
are obtained and its derivatives with respect to $q$ are plotted
against $q$ in Fig. \ref{3fss} (a) for $1$d with $\phi=2^{-6}$ and in
Fig.  \ref{3fss} (b) for $2$d with $\phi=2^{-8}$. It can be seen that
$\partial \sigma_s(q)/\partial q$ converges for large $q$ in both the
cases and the values of capacity dimension estimated as $D_s\approx
1.42$ and $\approx 2$ for $1$d and $2$d, respectively. It should be
noted here that for the given values of $\phi$, the distribution
functions consist of two scaling forms with exponents $\tau_{s1}$ and
$\tau_{s2}$. An attempt has been made to collapse the data by plotting
the scaled distribution $P(s,\phi)L^{D_s\tau_{s2}}$ against the scaled
variable $s/L^{D_s}$ in the respective insets.  For both the
dimensions, $\tau_{s2}$ is taken as $3/2$. Clearly, the collapse does
not work for the whole range of $s$. The deviation is more prominent
in the small avalanche size region where the avalanches follow usual
stochastic scaling of the respective dimensions. It is then important
to note that in the SWN regime FSS of the probability distributions
can not be verified for the stochastic sandpile model too because
multi-scaling feature of the same exists. A similar observation is
also reported by Benella {\em et.al.} \cite{benellaENTROPY17} in a
recent study of avalanching systems with long range
connectivity. Since FSS fails in the SWN regime, the $\phi$ dependent
coexistence scaling could be useful to verify the scaling form of the
probability distribution functions.

\subsection{Coexistence scaling} 
\begin{figure}[t]
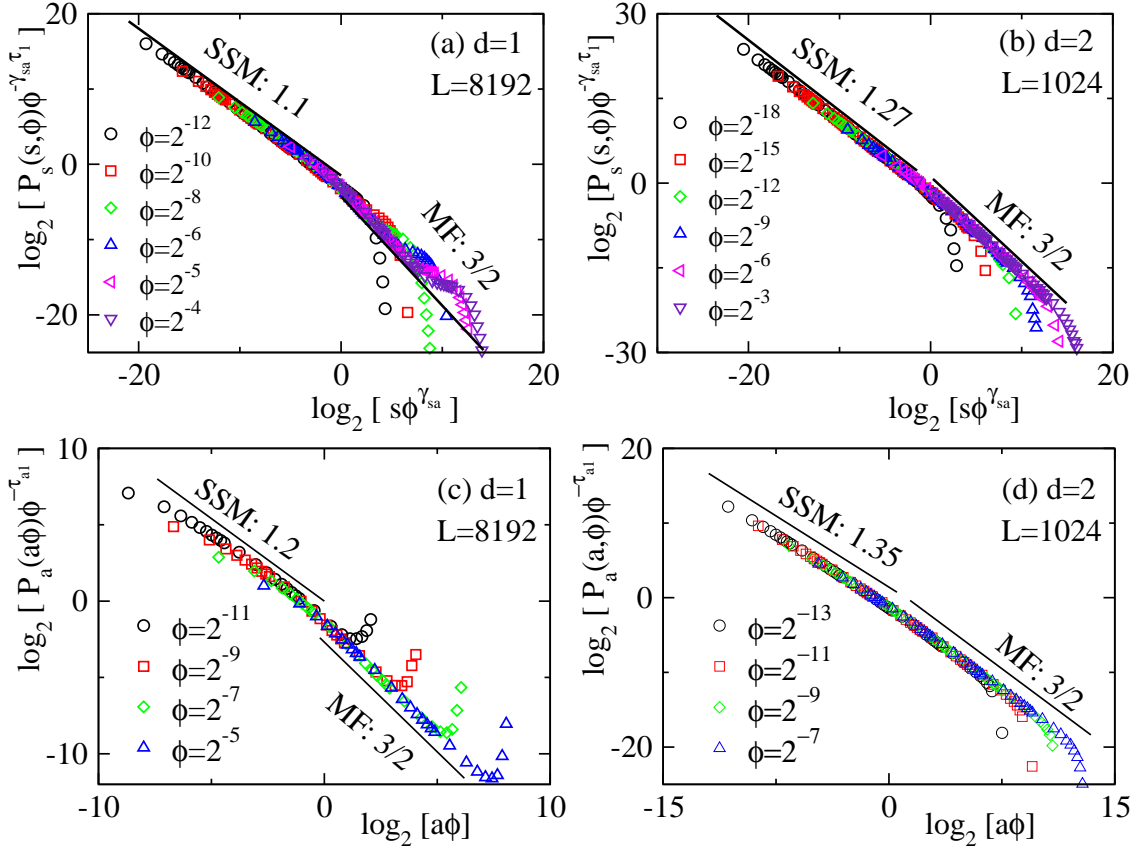

\centerline{\hfill
  \psfig{file=figure7ab.eps,width=0.90\textwidth}\hfill
} \centerline{\hfill
  \psfig{file=figure7cd.eps,width=0.90\textwidth}\hfill
}
\caption{ Plot of scaled size distributions against scaled variable
  for selected values of $\phi$ in (a) for $1$d and in (b) for
  $2$d. The scaled area distributions for several values of $\phi$ are
  shown in (c) for $1$d and in (d) for $2$d. The system size
  considered for $1$d is $L=8192$, and for $2$d is $L=1024$}
\label{3SC1}
\end{figure}
Verification of the scaling form given in Eq. (\ref{3scaling}) can be
performed for the two regions taking crossover value $x_c=x_1$ . At
$x_1(\phi)$ the values of $P_x(x,\phi)$ are same for both the regions.
Since $s_1\sim \phi^{-\gamma_{sa}}$, then one should have ${\sf
  f_s}(1) = \phi^{-(\tau_{s1}-\tau_{s2})\gamma_{sa}}{\sf
  g_s}(1)$. Following Ref. \cite{bhaumikPRE13}, the scaled size
distribution can be obtained as
\begin{eqnarray}
\label{3scaling2}
P_s(s,\phi)\phi^{-\gamma_{sa}\tau_{s1}} &=& \left \{ 
\begin{array}{ll} 
  \left(s\phi^{\gamma_{sa}}\right)^{-\tau_{s1}} {\sf
    f_s}(s\phi^{\gamma_{sa}}) & \mbox{for $s\leqslant s_1$}
  \\  \left(s\phi^{\gamma_{sa}}\right)^{-\tau_{s2}} {\sf
    f_s}(s\phi^{\gamma_{sa}}) &\mbox{for $s\geqslant s_1$ }\\
\end{array} \right. 
\end{eqnarray}
Similarly considering $a_1\sim s_1^{\gamma_{as}}\sim \phi^{-1}$
a scaled area distribution can be written as
\begin{eqnarray}
\label{3scaling2a}
P_a(a,\phi)\phi^{-\tau_{a1}} &=& \left \{ 
\begin{array}{ll} 
  \left(a\phi\right)^{-\tau_{a1}} {\sf f_a}(a\phi) & \mbox{for
    $a\leqslant a_1$} \\ \left(a\phi\right)^{-\tau_{a2}} {\sf
    f_a}(a\phi) &\mbox{for $a\geqslant a_1$ }\\
\end{array} \right. 
\end{eqnarray}
To verify the scaling forms given in Eqs. (\ref{3scaling2}) and
(\ref{3scaling2a}), the scaled distributions are plotted in
Fig. \ref{3SC1} against their respective scaled variable. Data of size
distributions are presented in Figs. \ref{3SC1}(a) and \ref{3SC1}(b)
for $1$d and $2$d respectively. It can be seen that a good data
collapse is obtained using $\gamma_{sa}=2$, $\tau_{s1}=1.11$ for $1$d
and using $\gamma_{sa}=1.26$, $\tau_{s1}=1.27$ for $2$d. Similarly
Figs. \ref{3SC1}(c) and \ref{3SC1}(d) represent the data of scaled
area distributions for $1$d and $2$d respectively. A reasonable
collapse are also observed taking $\tau_{a1}=1.2$ for $1$d and
$\tau_{a1}=1.35$ for $2$d. The straight lines with required slopes in
the respective regions are guide to the eye. It confirms the validity
of the proposed scaling function forms given in Eq. (\ref{3scaling}).

\begin{figure}[t]
\centerline{\hfill 
  \psfig{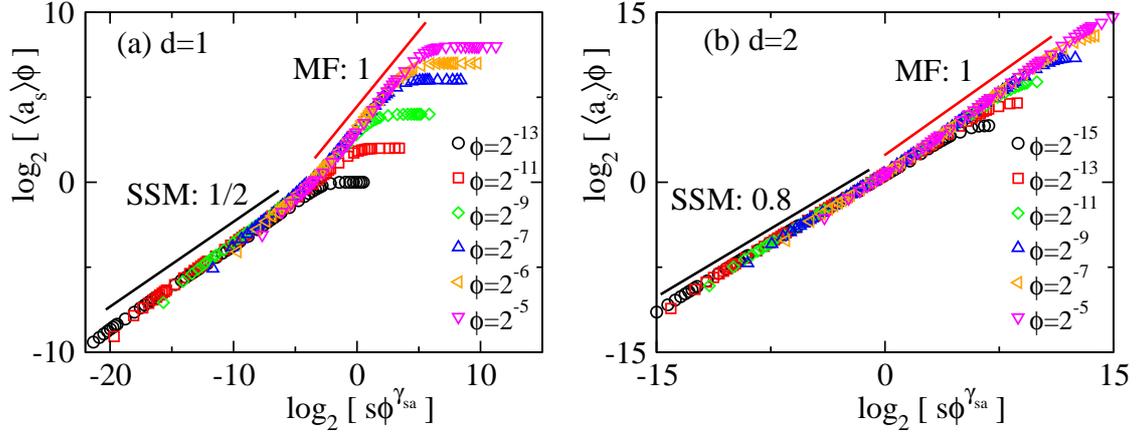}\hfill
 }
\caption{ Plot of scaled average area against scaled variable for
  selected values of $\phi$ in (a) for $1$d with $L=8192$ and (b) for
  $2$d with $L=1024$. Reasonable data collapse verify the scaling
  forms given in Eq. (\ref{3asscaling}). Straight lines with with
  required slope are guide to the eyes.}
\label{3ascoll}
\end{figure}

The coexistence scaling of the average height $\langle S_s\rangle$
around crossover size $s_1$ and $s_2$ has already been demonstrated in
previous section (Figs. \ref{3S}(c) and \ref{3S}(d)). Another
geometrical property of toppling surface, the average area $\langle
a_s \rangle$ of a given size $s$ also exhibits multiple scaling forms
in different regions of $s$. A coexistence scaling form of $\langle
a_s \rangle$ around the crossover size $s_1$ can be written as
\begin{equation}
\langle a_s \rangle = \left \{ \begin{array}{ll}
  s^{\gamma_{as1}}f_{as}\left( \frac{s}{s_1}\right) & \mbox{for $s
    \leqslant s_1$} \\ 
    s^{\gamma_{as2}}g_{as}\left( \frac{s}{s_1}\right) & \mbox{for $s
      \geqslant s_1$ }\\
\end{array} \right. 
\label{3asscaling}
\end{equation}
where, $f_{as}$ and $g_{as}$ are the scaling functions in respective
regions and $\gamma_{as1}$ and $\gamma_{as2}$ are the exponents in the
regime $s \leqslant s_1$ and $s \geqslant s_1$ respectively. Note that
at $s=s_1$ the value of $\langle a_s \rangle$ will be same for both
the regions and hence
$f_{as}(1)=s_1^{\gamma_{as2}-\gamma_{as1}}g_{as}(1)$. Considering
$\gamma_{sa}=1/\gamma_{as}$, Eq. (\ref{3asscaling}) can be written in
terms of single scaling function as
\begin{equation}
\langle a_s \rangle \phi = \left \{ \begin{array}{ll}
  (s\phi^{\gamma_{sa1}})^{\gamma_{as1}}f_{as}\left(s\phi^{\gamma_{sa1}}\right)
  & \mbox{for $s \leqslant s_1$}
  \\ (s\phi^{\gamma_{sa1}})^{\gamma_{as2}}f_{as}\left(
  s\phi^{\gamma_{sa1}}\right) & \mbox{for $s \geqslant s_1$ }\\
\end{array} \right. 
\label{3asscaling1}
\end{equation}
A plot of $\langle a_s \rangle \phi$ against $s\phi^{\gamma_{sa}}$ for
different values of $\phi$ should fall on a single curve. Satisfactory
collapse of date are observed as shown in Fig. \ref{3ascoll}(a) for
$1$d and in Fig. \ref{3ascoll}(b) for $2$d. Moreover, the scaling
function represents two different scaling behaviour with two different
exponents, $1/2$ and $1$ in $1$d and $0.8$ and $1$ in $2$d, as
indicated by straight lines with the respective slopes.

It is then important to notice that if a dynamical model like sandpile
is studied on SWN, multiple scaling forms of an event size will
coexist in the distribution of the event sizes. In other words, SWN
can be considered as a segregator of several scaling forms that appear
in the event size distribution.

\subsection{Sand transport: diffusivity and scaling} 
It is known in the literature that, the critical properties the
stochastic model on RL ($\phi=0$) governed by the diffusive nature of
the sand transport \cite{nakanishiPRE97,huynhJSM11}, i.e., $r\sim
\sqrt{t}$. However, as the SWN evolves to RN ($\phi=1$), it is
expected that the diffusive nature changes to super-diffusive one
i.e., $r\sim t$, as it is already observed for the deterministic model
\cite{bhaumikPRE13}. In order to verify the same, the average
avalanche size (time to diffuse) on a given system size $L$ (length to
diffuse) for a given $\phi$ is defined as
\begin{equation}
\langle s_\phi \rangle=\int s P_s(s,\phi)ds.
\end{equation}
\begin{figure}[t]
  \centerline{\hfill
    \psfig{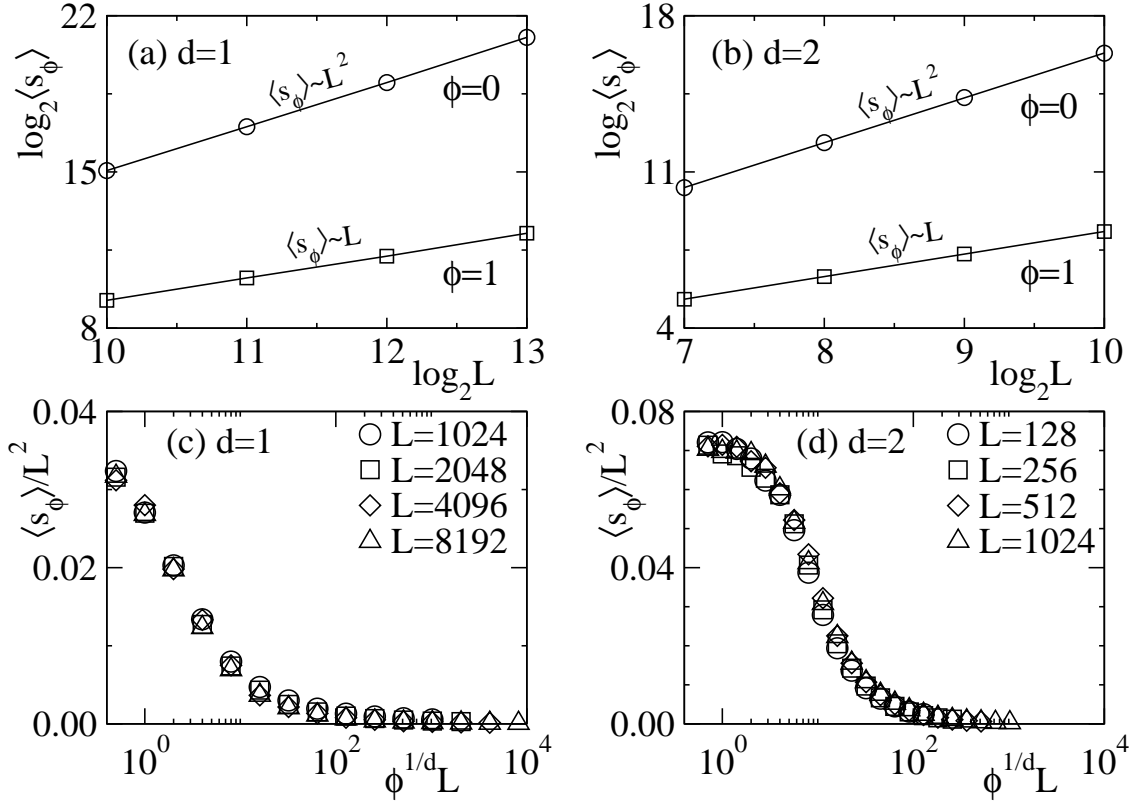}\hfill}
  \caption{ Plot of $\langle s_\phi \rangle$ against $L$ for $\phi=0$
    ($\Circle$) and $\phi=1$ ($\Box$) in (a) for $1$d and in (b) for
    $2$d. Plot of scaled average size $\langle s_\phi \rangle/L^2$
  against scaled variable $\phi^{1/d}L$ for (c) $d=1$ and (d)
  $d=2$. Different symbols correspond to different system sizes $L$.}
\label{3avgsL}
\end{figure}

At the extreme values of $\phi$, $\langle s_\phi \rangle$ is plotted
against system size $L$ in Figs. \ref{3avgsL}(a) and \ref{3avgsL}(b)
for $1$d and $2$d, respectively. It can be seen that in both the
dimensions, the diffusive behaviour of the model, $\langle s_\phi
\rangle \sim L^2$ holds for $\phi=0$, whereas the behaviour changes to
super-diffusive for other extreme value of $\phi$ i.e., $\langle
s_\phi \rangle \sim L$ for $\phi=1$. It is now important to check the
scaling behaviour of $\langle s_\phi \rangle$ with $\phi$ on SWNs. A
generalized scaling form of $\langle s_\phi \rangle$ can be written as
\begin{equation}
\langle s_\phi \rangle=L^2{\mathcal G}(\phi^{1/d}L)
\end{equation}
where $d$ is the space dimension. The behaviour of the scaling
function ${\mathcal G}(x)$ is given by,
\begin{eqnarray}
\label{3gscaling3}
 \mathcal{G}(x)&\propto& \left \{ \begin{array}{ll}
  \mbox{constant,} & x \ll 1,
  \\1/x,  & x \gg 1, \\
       \end{array} \right. 
\end{eqnarray}
In order to verify the above scaling form, $\langle s_\phi \rangle$ is
calculated for different system sizes over the full range of $\phi$
for both $1$d and $2$d. The scaled average size $\langle s_\phi
\rangle/L^2$ is plotted against the scaled variable $\phi^{1/d}L$ in
Figs. \ref{3avgsL}(c) and \ref{3avgsL}(d) for $1$d and $2$d
respectively. Taking the respective $d$ values, reasonable collapse of
data are observed in both the cases. Note that, for $2$d, the limiting
values of $\langle s_\phi \rangle/L^2$ as $\phi\to 0$ is found to be
$0.7$ which is consistence with the relations $2\langle s_\phi
\rangle=\langle n_\phi \rangle $ and $\langle n_\phi \rangle \approx
0.14L^2$, for $L\to\infty$ \cite{shiloPRE03}. However, for $1$d,
$\langle s_\phi \rangle/L^2$ is found to be $\approx 0.031$ as
$\phi\to 0$.

\section{Summary and Conclusion} 
A dissipative stochastic sandpile model (DSSM) is constructed and
studied on SWN both in $1$d and $2$d varying the shortcut density
$\phi$. As $\phi$ varies from $0$ to $1$, RL evolves to RN via a
series of SWNs. Since the critical behaviour of the stochastic model
on RL as well as on RN is known, emphasis is given in analyzing the
critical properties of the model on SWN regime ($2^{-12}<\phi
<0.1$). Several new geometrical quantities such as toppling surface
and its fragmentation, compactness and fluctuation in the fragment
size are defined and characterized as a function of avalanche size $s$
for a given $\phi$. The average height $\langle S_s\rangle$ and
average area $\langle a_s\rangle$ of the toppling surface are found to
have three regimes of avalanche size $s$, separated by two crossover
sizes $s_1$ and $s_2$ ($s_1<s_2$). Below $s_1$, the avalanches are
found to be compact as that are on RL. For $s_1<s<s_2$ regime, the
avalanches are fragmented in many sub-clusters that are connected by
long-ranged link of the network. Novel scaling forms of $s_1$ and
$s_2$ as well as that of $\langle S_s\rangle$ are developed and
numerically verified. Distributions of avalanche size $s$ and area $a$
are also found to exhibit two scaling forms about the crossover size
$s_1$. Below $s_1$ it is Manna scaling on RL whereas above $s_1$ it is
MF scaling on RN. Since two scaling forms appear simultaneously in the
SWN regime, the probability distributions of various avalanche
properties of DSSM do not follow the usual FSS in contrary to the fact
that stochastic sandpile model follow FSS on RL. Around the crossover
size a coexistence scaling form of the distributions and the
expectation value are developed and numerically verified. The sand
transport behaviour in DSSM is found to change from diffusive to super
diffusive nature as SWN evolve from RL to RN. A generalized scaling
form of diffusivity is identified that satisfactorily explains such
crossover behaviour.

\bigskip

\noindent{\bf Acknowledgments:} This work is partially supported by
DST, Government of India through project
No. SR/S2/CMP-61/2008. Availability of computational facility,
``Newton HPC'' under DST-FIST project (No. SR/FST/PSII-020/2009)
Government of India, of Department of Physics, IIT Guwahati is
gratefully acknowledged.

\bibliography{dssm_swn.bbl}
\end{document}